\documentclass[12pt]{article}
\usepackage{threeparttable}
 \usepackage{color}
\usepackage[margin=1in]{geometry}
\usepackage{amsmath,amsthm,amssymb,bm}
\usepackage{graphicx}
\usepackage{algorithm}
\usepackage{algorithmic}
\usepackage{multirow}
\usepackage{longtable}
\usepackage{rotating}
\usepackage{verbatim}

\usepackage{apacite}

\newcommand{\blue}[1]{{\textcolor{blue}{#1}}}

\begin{document}
\title{New Development of Bayesian Variable Selection Criteria for Spatial Point Process with Applications}
\author{Guanyu Hu \footnote{(to whom correspondence should be addressed) Email: {guanyu.hu@uconn.edu} Department of Statistics, University of Connecticut.}, Fred Huffer, and Ming-Hui Chen}
\date{}

\maketitle
\begin{abstract}
Selecting important spatial-dependent variables under the nonhomogeneous spatial Poisson process model is an important topic of great current interest.
 In this paper, we use the Deviance Information Criterion (DIC) and Logarithm of the Pseudo Marginal Likelihood (LPML) for Bayesian variable selection
 under the nonhomogeneous spatial Poisson process model. We further derive the new Monte Carlo estimation formula for LPML
in the spatial Poisson process setting. Extensive simulation studies are carried out to evaluate the empirical performance of the proposed criteria. The proposed methodology
is further applied to the analysis of two large data sets, the Earthquake Hazards Program of United States Geological Survey (USGS) earthquake data and the Forest of Barro Colorado Island (BCI) data.
 \end{abstract}
{\bf Keywords: BCI Data, DIC, LPML, Monte Carlo Estimation,  USGS Earthquake Data}

\section{Introduction}

There are three different types of spatial data, including point-reference data, areal data, and spatial point pattern data. Spatial point pattern data are the random locations of events in space
\cite{cressie1993statistics,banerjee2014hierarchical}. Spatial point patterns assume the randomness is associated with the locations of the points.
Spatial point process models have been developed for analyzing spatial point pattern data \cite{moller2003statistical,diggle2013statistical}.
Spatial point pattern data are routinely encountered in many different fields such as seismology, ecology, environmental science, and epidemiology. A common goal of these fields is to connect spatially dependent covariates to the occurrence of events of interest in space. Spatial point process regression models are well-suited for this goal.
Spatial Poisson point processes are widely used for the purpose of regression analysis due to the easy implementation and the attractive theoretical properties.

For regression problem, variable selection methods are widely studied in the statistical literature. The information criterion based variable selection methods such as \blue{Akaike information criterion} (AIC) \cite{akaike1974new} or \blue{Bayesian information criterion} (BIC) \cite{schwarz1978estimating} are easy to implement. Other than the information criterion based methods, the regularized regression methods such as ridge regression \cite{hoerl1970ridge}, LASSO \cite{tibshirani1996regression}, and elastic net \cite{zou2005regularization} are also important tools for variable selection. Within  the Bayesian framework,
the deviance information criterion (DIC) \cite{spiegelhalter2002bayesian} and the logarithm of the pseudo marginal likelihood (LPML) \cite{gelfand1994bayesian}
are two popular model selection assessment criteria. Priors based approaches (spike and slab prior \cite{ishwaran2005spike} and shrinkage prior \cite{polson2010shrink})
are also developed for Bayesian variable selection.

In the existing literature, the variable selection problem has been extensively studied for point-reference data or areal data (\cite{wang2009variable,reich2010bayesian,beale2010regression}).
However, for the spatial point process model, the literature on variable selection is relatively sparse.  \cite{thurman2014variable} proposed a regularized method that allows the selection of covariates at multiple pixel resolutions. \cite{yue2015variable} incorporated \blue{least absolute shrinkage and selection operator} (LASSO), adaptive LASSO and elastic net regularization methods into the generalized linear model framework to select important covariates for spatial point process models. \cite{thurman2015regularized} proposed a regularized estimating equation approach for the spatial point process model.
 \cite{leininger2017bayesian} proposed Bayesian model assessment methods using the predictive mean square error, empirical coverage, and ranked probability scores.
But these methods do not account for the tradeoff between the goodness of fit and the model complexity.

The major contribution of this paper is to develop a novel implementation of DIC and LPML under spatial point process models. Compared with point-reference data, the traditional Monte Carlo method for computing {Conditional Predictive Ordinate} CPO \cite{chen2012monte} is not applicable for spatial point pattern data. Therefore, we develop a new Monte Carlo method of CPO for spatial point pattern data. Compared with the regularized methods,
it is easy to carry out Bayesian inference using posterior samples based on our model selection approach.
Our simulation studies show the promising empirical performance of the proposed Bayesian estimate criteria. In addition, our proposed Bayesian approach reveals some interesting features of the
earthquake data sets. Furthermore, our proposed criteria overcome the limitations of regularized methods in selecting the important covariates and pixel resolution of the covariates simultaneously for BCI data.

The rest of this paper is organized as follows. In Section \ref{sec:bspp}, we review nonhomogeneous spatial Poisson process and provide the Bayesian formulation of the spatial Poisson process model.
The detailed development of DIC and LPML for the spatial Poisson process model is given in Section \ref{sec:model_assess}.
Simulation studies are presented in Section \ref{sec:simu}. In Section \ref{sec:real_data}, we carry out an in-depth analysis of the
two data sets,  the Earthquake Hazards Program of United States Geological Survey (USGS) earthquake data and Forest of Barro Colorado Island (BCI) data.
We conclude the paper with a brief discussion in Section \ref{sec:discuss}.

\section{Bayesian Spatial Point Process}\label{sec:bspp}

{A spatial point pattern is a data set $\bm{y}=(s_1,s_2,...,s_\ell)$ consists the locations $(s_1,s_2,...,s_\ell)$ of points that are observed in a bounded region $\mathcal{B}\subseteq \mathcal{R}^2$, which is a realization of spatial point process $\bm{Y}$. $N_{\bm{Y}}(A)=\sum_{i=1}^\ell 1(s_i\in A)$ is a counting process associated with the spatial point process $\bm{Y}$,
 which counts the number of points of $\bm{Y}$ for area $A\subseteq\mathcal{B}$.}
 For the process $\bm{Y}$, there are many parametric distributions for a finite set of count variables like Poisson processes, Gibbs processes, and Cox processes.
 In this paper, our focus is on Poisson processes. For the Poisson process $\bm{Y}$ over $\mathcal{B}$, which has the intensity function $\lambda(\bm{s})$, $N_{\bm{Y}}(A)\sim Po(\lambda(A))$, where $\lambda(A)=\int_A\lambda(s)ds$. In addition, if $A_1$ and $A_2$ are disjoint, then $N_{\bm{Y}}(A_1)$ and $N_{\bm{Y}}(A_2)$ are independent, where $A_1 \subseteq \mathcal{B}$ and $A_2 \subseteq \mathcal{B}$.
  Based on the properties of the Poisson process, it is easy to obtain $E(N_{\bm{Y}}(A))=\text{Var}(N_{\bm{Y}}(A))=\lambda(A)$. When $\lambda(s)=\lambda$, we have the constant intensity over the space $\mathcal{B}$ and in this special case, $\bm{Y}$ reduces to a homogeneous Poisson process (HPP). For a more general case, $\lambda(s)$ can be spatially varying, which leads to a nonhomogeneous Poisson process (NHPP). For the NHPP, the log-likelihood on $\mathcal{B}$ is given by
\begin{align}
	\ell=\sum_{i=1}^k \text{log}\lambda(s_i)-\int_{\mathcal{B}} \lambda(s) ds,
	\label{pploglike}
\end{align}
where $\lambda(s_i)$ is the intensity function for location $s_i$. To build up a spatially varying intensity,  $\lambda(s_i)$ is often assumed to take the following regression form:
\begin{align}
	\lambda(s_i)=\lambda_0\exp(\bm{\beta}\bm{Z}(s_i)),
	\label{intensity regression}
\end{align}
where $\lambda_0$ is the baseline intensity function, $\bm{\beta}$ is the vector of regression coefficient, and $\bm{Z}(s_i)$ is the vector of spatially varying covariates on location $s_i$.

For the model defined in \eqref{pploglike} and \eqref{intensity regression}, a gamma distribution for $\lambda_0$ is a conjugate prior. For the regression coefficients $\bm{\beta}$,
there are no such conjugate priors for the log-likelihood in \eqref{pploglike}. We specify a non-informative normal prior for $\bm{\beta}$.
Plugging \eqref{intensity regression} in \eqref{pploglike} and using the above priors for $\lambda_0$ and $\bm{\beta}$, we have
\begin{align}
	\begin{split}
		\ell&=\sum_{i=1}^k (\text{log}\lambda_0+\bm{\beta}\bm{Z}(s_i))-\int_\mathcal{B}\lambda_0\exp(\bm{\beta}\bm{Z}(s)) ds,\\
		\lambda_0 &\sim \text{G}(a_1,b_1),\\
		\bm{\beta} &\sim \text{MVN}(0,\sigma_0^2I),
	\end{split}
	\label{bnnpp}
\end{align}
where $l$ is the log-likelihood, and ``G",``MVN", and ``IG" are the shorthand of gamma distribution, multivariate normal distribution, and the inverse gamma distribution, respectively.
Thus, the posterior distribution of $\bm{\theta}=(\lambda_0,\beta_1,\cdots,\beta_p)$ is given as follow
\begin{align}
 \prod_{i=1}^n
    \lambda(s_i)\times \exp\left(-\int_\mathcal{B} \lambda(s)ds\right)\times \pi(\bm{\theta}),
\label{eq:posterior}
\end{align}
where $\pi(\bm{\theta})=\pi(\bm{\beta})\pi(\lambda_0)$ is the joint prior.
The analytical evaluation of the posterior distribution of $\bm{\theta}$ is not possible.
However, a Metropolis-Hasting algorithm within the Gibbs sampler can be developed to sample from the posterior distribution in \eqref{eq:posterior}.
The algorithm requires sampling the following parameters in turn from their respective full conditional distributions:
\begin{align}
\begin{split}
	f(\bm{\beta}|\cdot)&\propto L\times \prod_{k=1}^p \frac{1}{\sqrt{2\pi \sigma^2_0}}\exp(-\frac{\beta^2_k}{2\sigma^2_0}), \\
	f(\lambda_0|\cdot)&\propto L\times\frac{1}{\Gamma(a_1)b_1^{a_1}}{\lambda_0}^{a_1-1}\exp(-\frac{\lambda_0}{b_1}).
\end{split}
	\label{full_conditional1}
\end{align}
We choose $\sigma_0^2=100$ and $a_1=b_1=0.01$, while yields a non-informative joint prior for $\bm{\theta}$.
{To sample from the posterior distribution of $\bm{\theta}$ in
\eqref{eq:posterior}, an Metropolis--Hasting within Gibbs
algorithm is facilitated by R package \textsf{nimble}
\cite{de2017programming}. The loglikelihood function of the spatial Poisson point process model used in the MCMC iteration is directly defined using the \textsf{RW\_llFunction()} sampler.}


\section{Bayesian Criteria for Variable Selection}\label{sec:model_assess}

We first introduce two Bayesian model selection criteria, Deviance Information Criterion (DIC) and logarithm of the Pseudo-marginal likelihood (LPML).
The deviance information criterion is defined as
\begin{align}
	\text{DIC}=\text{Dev}(\bar{\bm{\theta}})+2p_D,
	\label{DIC}
\end{align}
where $\text{Dev}(\bar{\bm{\theta}})$ is the deviance function, $p_D=\overline{\text{Dev}}({\bm{\theta}}) - \text{Dev}(\bar{\bm{\theta}})$ is the effective number of model parameters,
$\bar{\bm{\theta}}$ is the posterior mean of parameters $\bm{\theta}$, and $\overline{\text{Dev}}(\bm{\theta})$ is the posterior mean of $\text{Dev}(\bm{\theta})$.

The LPML is defined as
\begin{align}
\label{eq:defLPML}
\text{LPML} = \sum_{i=1}^{n} \text{log}(\text{CPO}_i),
\end{align}
where $\text{CPO}_i$ is the conditional predictive ordinate (CPO) for the $i$-th subject. CPO is based on the leave-one-out-cross-validation. CPO estimates the probability of observing $y_i$ in the future
after having already observed $y_1,\cdots,y_{i-1},y_{i+1},\cdots,y_n$. The CPO for the $i$-th subject is defined as
\begin{align}
	\text{CPO}_i=f(y_i|\bm{y}_{-i}) \equiv \int f(y_i|\bm{\theta})\pi(\bm{\theta}|\bm{y}_{(-i)})d\bm{\theta},
	\label{cpo def}
\end{align}
where $\bm{y}_{-i}$ is $y_1,\cdots,y_{i-1},y_{i+1}\cdots,y_n$,
\begin{align}
	\pi(\bm{\theta}|\bm{y}_{-i})=\frac{\prod_{j\neq i}f(y_j|\bm{\theta})\pi(\bm{\theta})}{c(\bm{y}_{-i})},
\end{align}
and $c(\bm{y}_{-i})$ is the normalizing constant.
The $\text{CPO}_i$ in \eqref{cpo def} can be expressed as
\begin{align}
	\text{CPO}^{-1}_i=\frac{1}{\int\frac{1}{f(y_i|\bm{\theta})}\pi(\bm{\theta}|\bm{y})d\bm{\theta}}.
	\label{CPO1}
\end{align}
Using \eqref{CPO1}, a Monte Carlo estimate of  $\text{CPO}_i$ in \eqref{CPO1} is given by
\begin{align}
	\widehat{\text{CPO}}^{-1}_i=\frac{1}{B}\sum_{b=1}^B\frac{1}{f(y_i|\bm{\theta}_b)},
	\label{CPO monte}
\end{align}
where $\bm{\theta}_b$ is the $b$-th MCMC sample of $\bm{\theta}$ from $\pi(\bm{\theta}|\bm{y})$.

In the context of variable selection, we select a variable subset model which has the smallest DIC and the largest LPML.

\subsection{DIC for Spatial Point Process}
Since our main objective is to assess the fit of the spatial point pattern, we specify the following deviance function
\begin{align}
	\label{eq:deviance_spp}
	\text{Dev}\left(D,\bm{Z},\bm{\beta},\lambda_0\right)=-2\times\left(\sum_{i=1}^k \text{log}\lambda(s_i)-\int_{\mathcal{B}} \lambda(s) ds\right),
\end{align}
where $D=(\bm{s}_1,\bm{s}_2,\cdots,\bm{s}_k)$ is the observed points, $\lambda(s)=\lambda_0\exp(\beta_1Z_1(s)+\cdots+\beta_pZ_p(s))$ is the intensity function on location $\bm{s}$, and $Z_1(s),Z_2(s),\cdots,Z_p(s)$ are spatial covariates. Therefore, the DIC for the spatial point pattern is given  as follows
\begin{align}
	\label{eq:dic_spp}
	\text{DIC}=2E[\text{Dev}(D,\bm{Z},\bm{\beta},\lambda_0)]-\text{Dev}(D,\bm{Z},\hat{\bm{\beta}},\hat{\lambda}_0),
\end{align}
where $\hat{\bm{\beta}}$ and $\hat{\lambda}_0$ are the posterior means of $\bm{\beta}$ and $\lambda_0$ and
the expectation is taken with respect to the posterior distribution of $\bm{\beta}$ and $\lambda_0$.

\subsection{LPML for Spatial Point Processes}

In this section we propose a definition of LPML which is suitable for
general Poisson processes.  This definition is derived from the
definition of LPML in (\ref{eq:defLPML}) by a limiting argument.  We
begin by slightly altering the definition of $\text{CPO}_i$ given in
(\ref{cpo def}).  Let $f_0$ denote a fixed reference model for the data
$\bm{y}=(y_1, y_2,\ldots, y_n)$, and define
\begin{equation}
  \label{eq:altCPO}
  \text{CPO}_i = \frac{f(y_i | \bm{y}_{-i})}{f_0(y_i | \bm{y}_{-i})}\,.
\end{equation}
This definition is essentially equivalent to that in (\ref{cpo def})
since the reference model cancels out when comparisons are made
between models using $\text{LPML} = \sum_{i=1}^n \log(\text{CPO}_i)$.
We shall assume that the reference model $f_0$ involves no parameters
(or that any parameters are kept fixed at specified default values)
and that the observations $y_1, y_2, \ldots, y_n$ are independent.  In
this case $f_0(y_i | \bm{y}_{-i}) = f_0(y_i)$.  Repeating the argument
which leads to (\ref{CPO1}) then yields
\begin{equation}
  \label{eq:CPO1mod}
  \text{CPO}_i^{-1} =
\int \frac{f_0(y_i)}{f(y_i | \bm{\theta})}
\pi(\bm{\theta}|\bm{y})\,d\bm{\theta}.
\end{equation}

Suppose we have a full Bayesian model (such as that described in
(\ref{intensity regression}) and (\ref{bnnpp})) for a Poisson process
$\bm{Y}$ with intensity $\lambda_{\bm{\theta}}(\cdot)$ on a region
$\mathcal{B}\subset \mathcal{R}^2$.  We observe the realization
$Y = \{s_1, s_2, \ldots, s_k\}$.

Let $A_1, A_2, \cdots, A_n$ be a partition of $\mathcal{B}$, i.e., disjoint
subsets such that $\bigcup_{i=1}^n A_i = \mathcal{B}$.  Suppose we
were to make inference on $\bm{\theta}$ based only on the counts
$(N_{\bm{Y}}(A_1),N_{\bm{Y}}(A_2),\ldots,N_{\bm{Y}}(A_n))$ where
$N_{\bm{Y}}(A_i)=\sum_{j=1}^k 1(s_j\in A_i)$.  Conditional on
$\bm{\theta}$, these counts are independent Poisson random variables
which we take to be the values $y_1,y_2, \ldots, y_n$ making up the data
$\bm{y}$ in the generic notation of equations (\ref{eq:altCPO}) and
(\ref{eq:CPO1mod}).  For our reference model we use the Poisson
process with a constant intensity of one.  With these choices, we make
the following substitutions in (\ref{eq:CPO1mod}):
\begin{align*}
  f(y_i | \bm{\theta})
&= \frac{\lambda(A_i)^{N_{\bm{Y}}(A_i)}}{N_{\bm{Y}}(A_i)!} \exp\{-\lambda(A_i)\},\\
  f_0(y_i) &= \frac{|A_i|^{N_{\bm{Y}}(A_i)}}{N_{\bm{Y}}(A_i)!} \exp(-|A_i|),
\end{align*}
where $\lambda(A_i) = \int_{A_i} \lambda(u)\,du$ and
$|A_i| = \int_{A_i} du$.  (Here, to lighten the notation, we omit the
dependence of $\lambda$ on $\bm{\theta}$.)  Therefore
\begin{align*}
\frac{f_0(y_i)}{f(y_i| \bm{\theta})}
&= \left(\frac{\lambda(A_i)}{|A_i|}\right)^{-N_{\bm{Y}}(A_i)}
\exp(\lambda(A_i)-|A_i|),
\end{align*}
and we may write (\ref{eq:CPO1mod}) as
\begin{equation*}
  \text{CPO}_i^{-1} = E_n\left[\left(\frac{\lambda(A_i)}{|A_i|}\right)^{-N_{\bm{Y}}(A_i)}
\exp\{\lambda(A_i)-|A_i|\}\right],
\end{equation*}
where $E_n$ represents the expectation over the posterior distribution
of $\bm{\theta}$ given $(N_{\bm{Y}}(A_i),\ldots, N_{\bm{Y}}(A_n))$.

For the fixed realization of the Poisson process
$\bm{Y}= \{s_1, s_2, \ldots, s_k\}$, we consider (informally) taking a
limit as the partition $A_1, A_2, \ldots, A_n$ becomes progressively finer.
When the partition is sufficiently fine, no set $A_i$ will contain
more than one of the points $s_j$; exactly $k$ of the sets $A_i$ will
contain one point, and the others will contain zero points.  When the
set $A_i$ is sufficiently small, we expect
$e^{\lambda(A_i)-|A_i|} \approx 1 + (\lambda(A_i)-|A_i|)$.  Therefore,
when all the sets $|A_i|$ are sufficiently small, for those sets $A_i$
which contain zero points of $\bm{Y}$, we expect that
\begin{align*}
  \text{CPO}_i^{-1} &= E_n\left[ \exp\{\lambda(A_i)-|A_i|\}\right] \approx 1+ E_n(\lambda(A_i)-|A_i|),
\end{align*}
and {based on the first order of Taylor expansion of $\log(1+x)$, we have}
\begin{equation}
\label{zero points}
\log \text{CPO}_i \approx - E_n(\lambda(A_i)-|A_i|).
\end{equation}
If a set $A_i$ is sufficiently small and contains a single point $s_j$
of $\bm{Y}$, then (assuming the intensity $\lambda(u)$
is a continuous function of $u$ for all $\bm{\theta})$  {and $A_i$ is smaller enough such that have negligible area}) we expect
$\lambda(A_i)/|A_i| \approx \lambda(s_j)$ and therefore
\begin{align*}
  \text{CPO}_i^{-1}
&= E_n\left[\left(\frac{\lambda(A_i)}{|A_i|}\right)^{-1} e^{\lambda(A_i)-|A_i|}\right]
\approx E_n\lambda(s_j)^{-1}(1+(\lambda(A_i)-|A_i|)) \approx E_n\left[\lambda(s_j)^{-1}\right],
\end{align*}
and
\begin{align}
\label{one point}
\log \text{CPO}_i &\approx  \log \left(E_n\left[\lambda(s_j)^{-1}\right]\right)^{-1}.
\end{align}
Combining (\ref{zero points}) and (\ref{one point}), we obtain
\begin{align*}
  \text{LPML}_n &= \sum_{i=1}^n \log \text{CPO}_i
= \sum_{i:N_{\bm{Y}}(A_i)=1} \log \text{CPO}_i + \sum_{i:N_{\bm{Y}}(A_i)=0} \log \text{CPO}_i\\
&\approx \sum_{j=1}^k \log \left(E_n\left[\lambda(s_j)^{-1}\right]\right)^{-1}
- \sum_{i:N_{\bm{Y}}(A_i)=0} E_n(\lambda(A_i)-|A_i|)\\
&\approx \sum_{j=1}^k \log \left(E_n\left[\lambda(s_j)^{-1}\right]\right)^{-1}
- \int_{\mathcal{B}} E_n[\lambda(u)]\,du + |\mathcal{B}|.
\end{align*}

As the partition gets progressively finer, the posterior conditional
on the counts $(N_{\bm{Y}}(A_1),$
$N_{\bm{Y}}(A_2), \ldots,N_{\bm{Y}}(A_n))$ converges to
the posterior conditional on the complete data $\bm{Y}$ \cite{waagepetersen2004convergence}, and the
various approximations we had made above become exact in the limit.
We obtain
\begin{equation*}
\lim_{n \rightarrow \infty} \text{LPML}_n = \sum_{j=1}^k \log \left(E\left[\lambda(s_j)^{-1}\right]\right)^{-1}
- \int_{\mathcal{B}} E[\lambda(u)]\,du + |\mathcal{B}|,
\end{equation*}
where $E$ denotes the posterior expectation.  Dropping the constant term
$|\mathcal{B}|$, we take the remainder of this expression as our
definition of the LPML for Poisson processes:
\begin{equation}
\text{LPML} = \sum_{j=1}^k \log \left(E\left[\lambda(s_j)^{-1}\right]\right)^{-1}
- \int_{\mathcal{B}} E[\lambda(u)]\,du,
\end{equation}
where $\bm{Y}= \{s_1, s_2, \ldots, s_k\}$ and $E$ denotes the posterior
expectation.  There is a natural Monte Carlo estimate of the LPML
given by
\begin{equation}
  \widehat{\text{LPML}}
= \sum_{j=1}^k\log \widetilde{\lambda}(s_j)-\int_{\mathcal{B}}\overline{\lambda}(u)\,du,
\end{equation}
where
$\widetilde{\lambda}(s_j)=(\frac{1}{B}\sum_{b=1}^B\lambda(s_j|\bm{\theta_b})^{-1})^{-1}$,
$\overline{\lambda}(u)=\frac{1}{B}\sum_{b=1}^B
\lambda(u|\bm{\theta_b}) $, and $\{\theta_1,\theta_2,\cdots,\theta_B\}$ is a sample
from the posterior.

\section{Simulation Study}\label{sec:simu}

In this section, we have four scenarios to generate covariates. The intensity function of data generation model (DGM) in scenario 1 is $\lambda(s)=\lambda_0\exp(\beta_1\times Z_1(s)+\beta_2 \times Z_2(s)+\beta_3\times Z_1(s)Z_2(s))$, where $Z_1(s)$ and $Z_2(s)$ are $x$-coordinate and $y$-coordinate of location $s$. In this simulation we choose $\beta_1=2$, $\beta_2=0$,
$\beta_3=1$, and $\lambda_0=30$ and generate data on $[0,1]\times[0,1]$. We generate 100 data sets under this setting, and the average number of points in this scenario is around 130.  {In this simulation study, we compare seven different models. We give $N(0,10^2)$ prior on $\beta$s and $G(1,1)$ prior on $\lambda_0$. For each replicates, 20,000 MCMC samples are drawn and last 10,000 samples are kept for Bayesian inference.  The full regression model includes x-coordinate, y-coordinate and their cross effects.}
 {The average DIC and LPML values and the selection percentage and model informations are shown in Table \ref{table_scenario_3}}. This table indicates that (i) the true model has the smallest average DIC value and the largest average LPML value; (ii) the true model also has the largest selection percentages under both DIC and LPML; and (iii) Model 4 has the second smallest DIC value, the second largest LPML value,
and the second largest selection percentages. A partial explanation for (iii) is that both $Z_1(s)$ and $Z_2(s)$ are correlated with the interaction term  $ Z_1(s)Z_2(s)$.

\begin{table}[h]
\center
\caption{Selection Results of Scenario 1}
	\begin{tabular}{ccccc}
		\hline
		Model&Average DIC& Average LPML&DIC Selection \% & LPML Selection \%\\
		\hline
		Model 1($\beta_1$)&-1147.300&573.639&12&12\\
		Model 2 ($\beta_2$)&-1093.641&546.810&0&0\\
		Model 3 ($\beta_3$)&-1133.599&566.790&0&0\\
		Model 4 ($\beta_1,\beta_2$)&-1151.719&575.840&24&25\\
		DGM ($\beta_1,\beta_3$)&-1152.609&576.284&53&52\\
		Model 5 ($\beta_2,\beta_3$)&-1145.189&572.573&5&5\\
		Model 6 ($\beta_1,\beta_2,\beta_3$)&-1151.656&575.788&6&6\\
		\hline
	\end{tabular}
	\footnotetext[1]{True Model}
	
	\label{table_scenario_3}
\end{table}

\begin{figure}[h]
	\center
		\caption{DIC and LPML Differences of Scenario 1}
	\includegraphics[width =5 in]{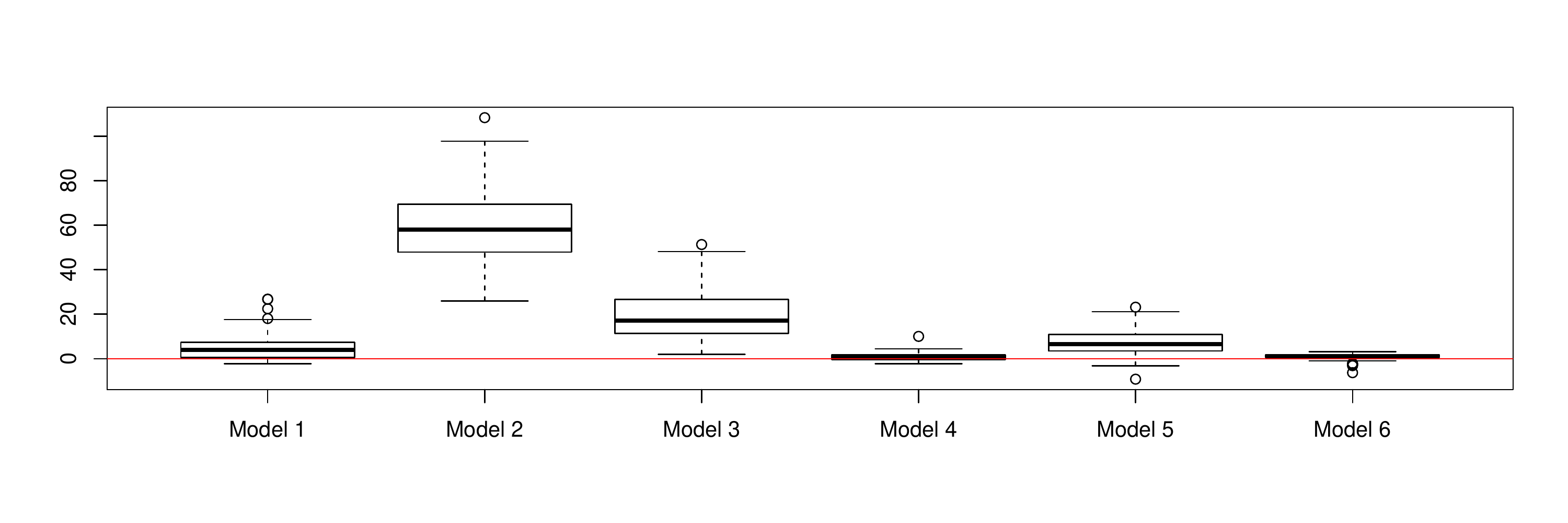}\\
	\includegraphics[width =5 in]{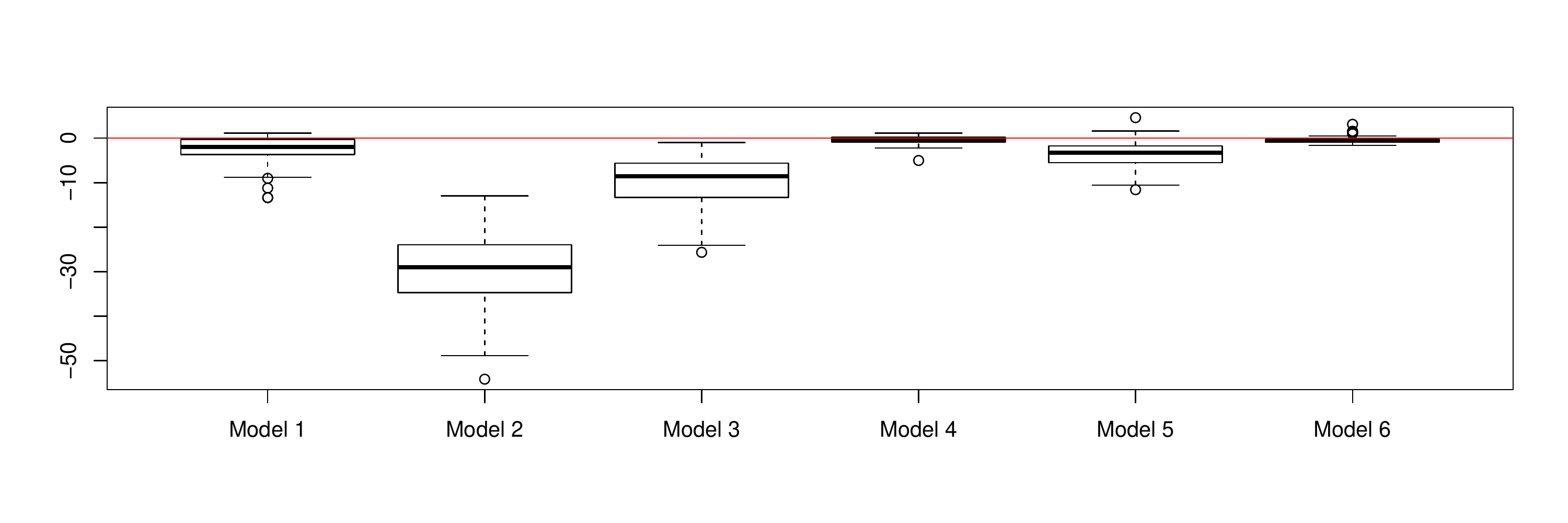}
	\label{figure_scenario_3}
\end{figure}

Furthermore, we also show the boxplots of the DIC difference and the LPML difference between the true model and each of the candidate models in Figure \ref{figure_scenario_3}.  {The medians and the inter quartile ranges (IQRs) of the DIC differences and the LPML differences are show in Table \ref{table:median_IQR_DIC_LPML}}
\begin{table}[h]
\center
\caption{Median and inter quartile ranges (IQRs) of the DIC and the LPML differences for Scenario 1}
\begin{tabular}{lll}
\hline
    & DIC & LPML \\
    \hline
Model 1 &     4.033 (0.517,7.339)   &   -2.009 (-0.253, -3.654)     \\
Model 2 &    57.98 (48.03,69.35)   &  -28.97 (-24.00, -34.66)      \\
Model 3 &   17.07 (11.60,26.40)    &  -8.520 (-5.789, -13.18)     \\
Model 4 &     0.658 ( -0.378,1.798)   &  -0.339 (0.197, -0.887)     \\
Model 5 &      6.541 (3.48,10.95)  &  -3.281 (-1.735, -5.481)     \\
Model 6 &   1.244 (0.576,1.771)    &   -0.635 (-0.298, -0.908)\\
\hline    
\end{tabular}
\label{table:median_IQR_DIC_LPML}
\end{table}
 Except for Models 4  the first quartiles of DIC differences are above 0 while the third quartiles of the LPML difference are below 0, which indicate that the true model outperform
  models 1, 2, 3, 5, and 6 according to the  DIC and LPML values. However, the medians of the DIC and LPML differences are close to 0 for both Models 4 since
  both $Z_1(s)$ and $Z_2(s)$ are correlated with the interaction term  $ Z_1(s)Z_2(s)$.

 {Furthermore, the intensity of DGM is $\lambda(s)=\lambda_0\exp(\beta_1\times Z^2_1(s))$, where $Z_1(s)$ is the x-coordinate. We choose $\lambda_0=50$ and $\beta_1=4$. We generate data on $[0,1]\times[0,1]$ and the average number of points over 100 replicates is 412. We compare the data generation model with three different models: Model 1: $\lambda(s)=\lambda_0\exp(\beta_1\times Z_1(s))$; Model 2: $\lambda(s)=\lambda_0\exp(\beta_2\times Z_2(s))$, where $Z_2(s)$ is the y-coordinate; Model 3:  $\lambda(s)=\lambda_0\exp(\beta_1\times Z_1(s)+\beta_2\times Z_2(s))$. We have the same prior distributions and same number of MCMC samples with scenario 1}

 {The average DIC and LPML values and the selection percentage and model informations are shown in Table \ref{table_scenario_5} }

\begin{table}[h]
\center
\caption{Selection Results of Scenario 2}
	\begin{tabular}{ccccc}
		\hline
		Model&Average DIC& Average LPML&DIC Selection \% & LPML Selection \%\\
		\hline
		DGM&-4702.209&2351.101&94&94\\
		Model 1 &-4690.384&2345.186&6&6\\
		Model 2 &-4145.1794&2072.586&0&0\\
		Model 3 &-4689.554&2344.769&0&0\\
		\hline
	\end{tabular}
	\footnotetext[1]{True Model}
	
	\label{table_scenario_5}
\end{table}

 {Furthermore, the medians and the inter quartile ranges (IQRs) of the DIC differences and the LPML differences are show in Table \ref{table:median_IQR_DIC_LPML2}}
\begin{table}[h]
\center
\caption{Median and inter quartile ranges (IQRs) of the DIC and the LPML differences for Scenario 2}
\begin{tabular}{lll}
\hline
    & DIC & LPML \\
    \hline
Model 1 &     10.729 (6.676,17.11)   &   -5.365 (-8.562, -3.335)       \\
Model 2 &    551.8 (531.1,584.3)   &  -275.9 (-292.2, -1265.5)      \\
Model 3 &   12.16 (7.26,17.83)    &  -6.079 (-8,941,-3.628)     \\
\hline    
\end{tabular}
\label{table:median_IQR_DIC_LPML2}
\end{table}

 {From the results in Table \ref{table_scenario_5} and Table \ref{table:median_IQR_DIC_LPML2}, our proposed criteria can effectively select the true model and the difference of DIC and LPML between true model and candidate models is significant.}

 {The intensity function of DGM in scenario 3 is $\lambda(s)=\exp(\beta_1\times Z_1(s)+\beta_2\times Z_2(s))$. $\bm{Z}_i,i=1,\cdots,4$ are generated from Gaussian random fields based on R package \textbf{RandomFields}. Our choice of Gaussian random fields with mean 1 and the covariance with exponential kernel  model with variance 1, scale parameter 1 and nugget variance 0.2.} In this simulation we choose $\beta_1=2$, $\beta_2=1$, and $\lambda_0=1$ and generate data on $[0,1]\times[0,1]$.  {We have $100\times 100$ grids on $[0,1]\times[0,1]$. Based on intensity surface on $[0,1]\times[0,1]$, we independently generate locations in each grids. }
We generate 100 data sets under this setting, and the average number of points in each data set is 544.  {We have the same prior distributions and the same number of MCMC samples with previous simulation.}
The average DIC and LPML values and the selection percentage are shown in Table \ref{table_scenario_4}.
From this table, we see that the true model has the smallest DIC value, the largest LPML value, and the highest selection percentages according to both DIC and LPML.
\begin{table}[h]
\center
\caption{Selection Results of Scenario 3}
	\begin{tabular}{ccccc}
		\hline
		Model &Average DIC& Average LPML&DIC Selection \%& LPML Selection \%\\
		\hline
		Model 1 ($\beta_1$)&-8877.358&4438.677&0&0\\
		Model 2 ($\beta_2$)&-6402.960&3201.478&0&0\\
		Model 3 ($\beta_3$)&-5590.344&2795.170&0&0\\
		Model 4 ($\beta_4$)&-5521.848&2760.921&0&0\\
		DGM ($\beta_1,\beta_2$)&-9303.699&4651.845&74&74\\
		Model 5 ($\beta_1,\beta_3$)&-8890.624&4445.308&0&0\\
		Model 6 ($\beta_1,\beta_4$)&-8879.830&4439.910&0&0\\
		Model 7 ($\beta_2,\beta_3$)&-6439.253&3219.623&0&0\\
		Model 8 ($\beta_2,\beta_4$)&-6409.395&3204.694&0&0\\
		Model 9 ($\beta_3,\beta_4$)&-5577.677&2788.834&0&0\\
		Model 10 ($\beta_1,\beta_2,\beta_3$)&-9302.598&4651.292&15&15\\
		Model 11 ($\beta_1,\beta_2,\beta_4$)&-9302.500&4651.243&9&9\\
		Model 12 ($\beta_1,\beta_3,\beta_4$)&-8891.348&4445.667&0&0\\
		Model 13 ($\beta_2,\beta_3,\beta_4$)&-6443.625&3221.808&0&0\\
		Model 14 ($\beta_1,\beta_2,\beta_3,\beta_4$)&-9301.493&4650.738&2&2\\
		\hline
	\end{tabular}
	
	\label{table_scenario_4}
\end{table}

\begin{figure}[H]
	\center
	\caption{DIC and LPML Differences of Scenario 3}
	\includegraphics[width =5 in]{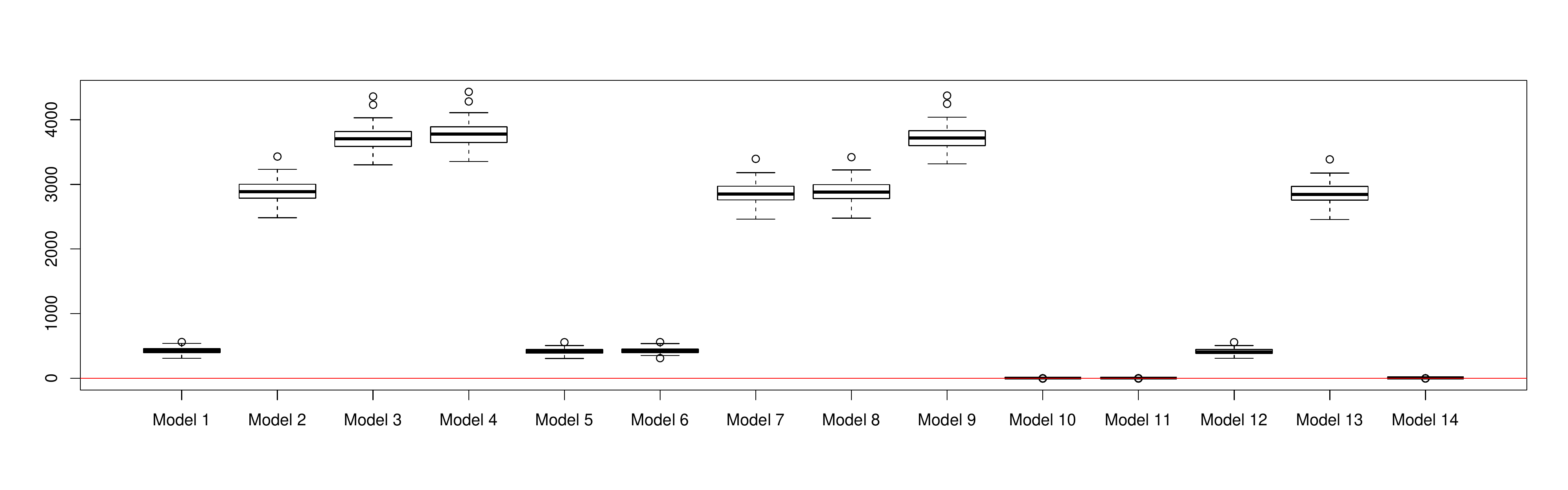}\\
	\includegraphics[width =5 in]{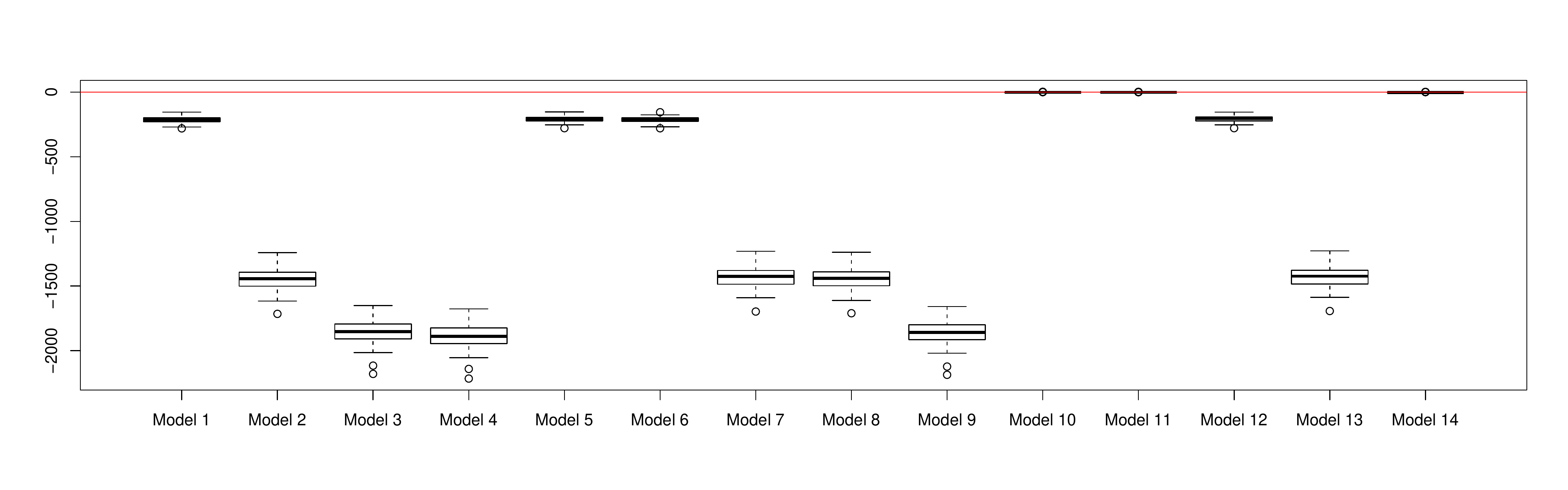}
	
	\label{figure_scenario_4}
\end{figure}

Furthermore, we also show the boxplots of the DIC difference and the LPML difference between the true model and each of the candidate models in Figure \ref{figure_scenario_4}. The medians and the inter quartile ranges (IQRs) of the DIC differences in the top panel of Figure \ref{figure_scenario_4} are 1.608 (0.651, 2.012), 1.698 (0.826, 1.968), and 2.944 (1.492, 3.501) for Models 10, 11 and 14, respectively;
and  the medians and the inter quartile ranges (IQRs) of the LPML differences in the bottom panel of Figure \ref{figure_scenario_4} are -0.805 (-0.412, -0.985), -0.849 (-0.412, -0.985), and -1.475 (-0.754, -1.758) for Models 10, 11 and 14, respectively.
Unlike scenario 3, the whole boxes of the DIC differences are above 0 while the boxes of the LPML differences are below 0.
Thus, there is an overwhelming evidence in favor of the true model according both DIC and LPML.


 {Finally, the intensity of DGM of scenario 4 is $\lambda(s)=\exp(\beta_1\times Z_1(s)+\beta_2\times Z_2(s)+W(s))$, $\bm{Z}_i,i=1,\cdots,3$ are generated from uniform distribution $U(0,1)$. $W(s)$ is  generated from Gaussian random with mean 0 and the covariance with exponential kernel  model with variance 1, scale parameter 1 and nugget variance 0.2, which is unobserved in this simulation. In this simulation we choose $\beta_1=4$, $\beta_2=4$, and $\lambda_0=1$ and generate data on $[0,1]\times[0,1]$. We generate 100 data sets under this setting, and the average number of points in each data set is 182. The full model is this simulation is $\lambda(s)=\exp(\beta_1\times Z_1(s)+\beta_2\times Z_2(s)+\beta_3\times Z_3(s))$. We have $100\times 100$ grids on $[0,1]\times[0,1]$. Based on intensity surface on $[0,1]\times[0,1]$, we independently generate locations in each grids. We have the same prior distributions and same number of MCMC samples with previous scenarios. The average DIC and LPML values and the selection percentage are shown in Table \ref{table_scenario_4}.
From this table, we see that the true model has the smallest DIC value, the largest LPML value, and the highest selection percentages according to both DIC and LPML. From this table, we see that the true model has the smallest DIC value, the largest LPML value, and the highest selection percentages according to both DIC and LPML.}
\begin{table}[h]
\center
\caption{Selection Results of Scenario 4}
	\begin{tabular}{ccccc}
		\hline
		Model &Average DIC& Average LPML&DIC Selection \%& LPML Selection \%\\
		\hline
		Model 1 ($\beta_1$)&-1694.704&847.344&0&0\\
		Model 2 ($\beta_2$)&-1696.554&848.269&0&0\\
		Model 3 ($\beta_3$)&-1459.131&729.560&0&0\\
		DGM ($\beta_1,\beta_2$)&-1873.874&936.921&88&88\\
		Model 4 ($\beta_1,\beta_3$)&-1693.393&846.683&0&0\\
		Model 5 ($\beta_2,\beta_3$)&-1695.379&847.676&0&0\\
		Model 6 ($\beta_1,\beta_2,\beta_3$)&-1872.952&936.454&12&12\\
		\hline
	\end{tabular}
	
	\label{table_scenario_4}
\end{table}

\section{Real Data Analysis}\label{sec:real_data}


\subsection{USGS Earthquake Data}

\begin{figure}
	\caption{Map of Locations of the Earthquake}
	\center
	\includegraphics[width= 3 in]{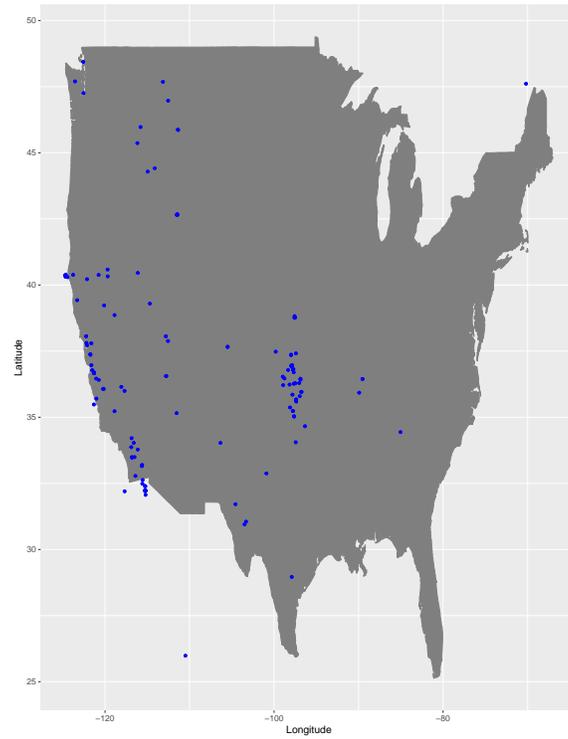}
	\label{fig:earthquake_map}
\end{figure}

\begin{figure}
	\center
	\caption{Map of Locations of the Earthquake in Unit Square}
	\includegraphics[width= 3 in]{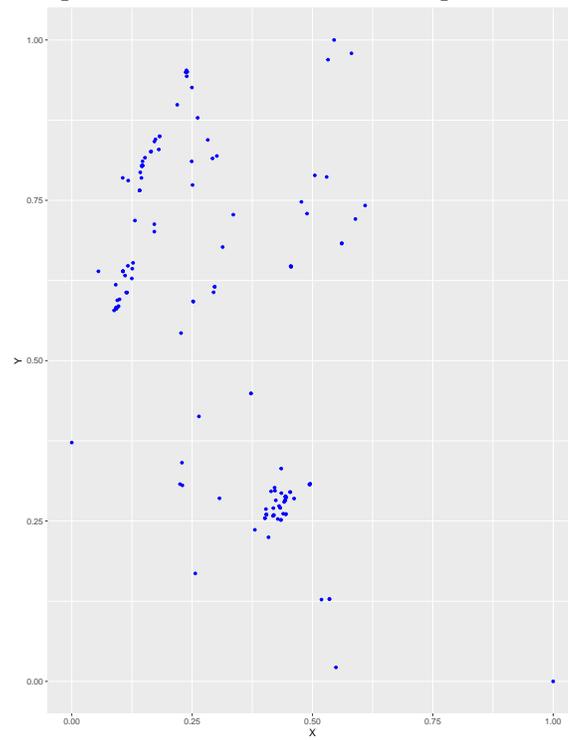}
	\label{fig:earthquake_square}
\end{figure}

The earthquake data from USGS, the Earthquake Hazards Program of United States Geological Survey (USGS), can be accessed via
  {https://earthquake.usgs.gov/earthquakes/}.
 The data set we consider is composed of conterminous U.S. earthquakes which have magnitude over 2.5 from 10-30-2018 to 11-27-2018. 
 From USGS, conterminous U.S. refers to a rectangular region including the lower 48 states and surrounding areas which are outside the Conterminous U.S..
The total number of earthquakes is 155. The map of the locations of earthquakes is shown in Figure \ref{fig:earthquake_map}.
In order to properly examine the relationship of these locations,  {we transform the latitude and the longitude of the earthquakes to UTM (Universal Transverse Mercator) coordinate system and then scale UTM coordinate system to a $[0,1]\times [0,1]$ square.} 
The locations of earthquakes in a unit square are shown in Figure \ref{fig:earthquake_square}.
The spatial covariates used in this analysis include $x$-coordinates which are transferred by longitudes, $y$-coordinates which are transferred by latitudes, and the distance to New Madrid Seismic Zone \cite{tuttle2005evidence}  {which has coordinates (36.58 N, 89.58 W). Similarly, we transform the latitude and the longitude of the New Madrid Seismic Zone to UTM system and scale it to the same unit square.} New Madrid Seismic Zone is a major seismic zone and a prolific source of intraplate earthquakes in the southern and midwestern United States. 
The full spatial intensity model for the spatial point process is given by
\begin{align}
	\label{eq:spatial_pp_iten_earthquake}
	\lambda(s)=\lambda_0\exp(\beta_1Z_1(s)+\beta_2Z_2(s)+\beta_3Z_3(s)),
\end{align}
where $Z_1(s)$ is $x$-coordinate, $Z_2(s)$ is $y$-coordinate, and $Z_3(s)$ is the distance to the New Madrid Seismic Zone.  
Including the homogeneous spatial Poisson process model, we have 8 candidate models. With the thinning interval to be 10, 2,000 samples are kept for calculation after a burn-in of 30,000 samples 
using NIMBLE in R for each models. {The prior for $\beta$s is $N(0,10^2)$ and the prior for $\lambda_0$ is $G(1,1)$.} The DIC and LPML values of these 8 candidate models are given in Table \ref{table:earthquake_DIC_LPML}.
\begin{table}[h]
\caption{DIC and LPML of Earthquake Data}
\label{table:earthquake_DIC_LPML}
\center
	\begin{tabular}{cccccc}
	\hline
		Model& DIC&LPML&Model& DIC&LPML\\
		\hline
		$\beta_1,\beta_2,\beta_3$&-1268.687&634.329&$\beta_1,\beta_2$&-1261.741&630.864\\
		$\beta_1,\beta_3$&-1239.627&619.8027&$\beta_2,\beta_3$&-1218.654&609.317\\
		$\beta_1$&-1179.762&589.878&$\beta_2$&-1219.484&609.7372\\
		$\beta_3$&-1206.92&603.452&Homogenous&-1192.528&596.262\\
		\hline
	\end{tabular}
\end{table}
The DIC and LPML values in Table \ref{table:earthquake_DIC_LPML} suggest that the model with $\beta_1,\beta_2,\beta_3$ is the best.  The posterior means, the posterior standard deviations and the 95\% 
   highest posterior density (HPD) intervals \cite{chen1999monte} under the best model are reported in Table \ref{table:earthquke_estimation}.
\begin{table}[h]
\caption{Best Model Parameter Estimation of Earthquake Data}
\label{table:earthquke_estimation}
\center
	\begin{tabular}{cccc}
	\hline
		Parameter& Posterior Mean & Posterior Standard Deviation & HPD interval\\
		\hline
		$\beta_1$&-1.975&0.352&$(-2.699,-1.343)$\\
		$\beta_2$&5.721&1.161&$(3.621,7.884)$\\
		$\beta_3$&-2.744&1.213&$(-4.909,-0.512)$\\
		$\lambda_0$&37.850&5.491&$(28.314,49.394)$\\
	\hline	
	\end{tabular}
\end{table}
From  Table \ref{table:earthquke_estimation}, we see that the occurrence of earthquake has a higher intensity around the New Madrid Seismic Zone and the southwestern area of U.S., which is consistent with the findings in the literature \cite{fuller1988new}.

\subsection{Forest of Barro Colorado Island (BCI)}

\begin{figure}
\label{tree_map}
	\caption{Locations Map of \textit{B.pendula}}
\center
	\includegraphics[width=4 in]{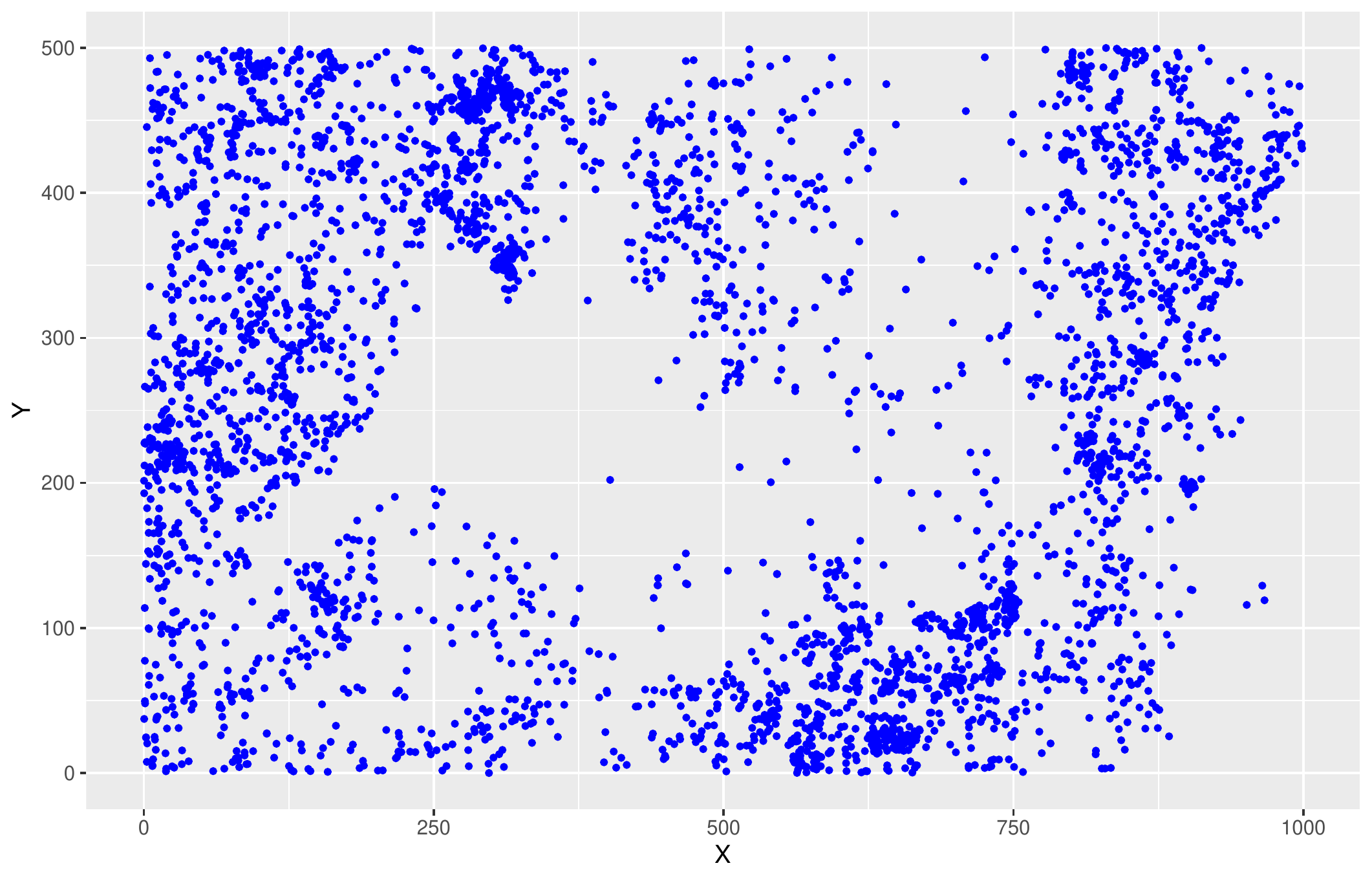}

\end{figure}

BCI data have be analyzed in the literature \cite{leininger2017bayesian,thurman2015regularized,yue2015variable,thurman2014variable}. The data are  {obtained} from a long-term ecological monitoring program, which contain 400,000 individual trees  since the 1980s at the Barro Colorado Island (BCI) in central Panama \cite{hubbell1999light,condit1998tropical,condit2012dataset}. We choose the same species like \cite{thurman2014variable} in our data analysis. We model the intensity of the \textit{B.pendula} trees as a log-linear function of the incidences of six other tree species (\textit{Eugenia nesiotica, Eugenia oerstediana, Piper cordulatum, Protium panamense, Sorocea affinis,} and \textit{Talisia croatii}). The map of the locations of \textit{B.pendula} tress mentioned above is shown in Figure \ref{tree_map}.

We choose three different pixel resolutions, $5 m \times 5 m$, $10 m \times 10 m$, and $20 m \times 20 m$ for the incidences of six other tree species (6 spatial covariates). 
The full spatial intensity model for the spatial point process is given by
\begin{align}
	\label{eq:spatial_pp_iten_earthquake}
	\lambda(s)=\lambda_0\exp(\beta_1 Z_1(s)+\cdots+\beta_6 Z_6(s)),
\end{align}
where $Z_i(s)$ is the $i$-th specie's pixel value on location $s$. Including the homogeneous spatial Poisson process model, we have 64 candidate models. 8,000 samples are kept for calculation after a burn-in of 10,000 samples using NIMBLE in R for each model. {The prior for $\beta$s is $N(0,10^2)$ and the prior for $\lambda_0$ is $G(1,1)$.}  The DIC and the LPML values of the best models for the three pixel resolutions are given in Table \ref{table:pixel_comparision123}.
\begin{table}[h]

\center
\caption{Model Comparison for Different Pixel Resolutions}
	\begin{tabular}{cccc}
	\hline
	Pixel Resolution& Best Model & DIC&LPML\\
	\hline
	 $5 m \times 5 m$&$(\beta_4,\beta_5)$&-57521.48&28760.74\\
	  $10 m \times 10 m$&$(\beta_1,\beta_3,\beta_4,\beta_5)$&-58255.15&29127.57\\
	   $20 m \times 20 m$&$(\beta_3,\beta_4,\beta_5)$&-58188.46&29094.22\\
	   \hline	  	
	\end{tabular}
	\label{table:pixel_comparision123}
\end{table}

\begin{figure}[h]
 \center
 \caption{$10 m \times 10 m$ Pixel Resolution Covariates Map (Red indicate larger values)}
 \begin{tabular}{ccc}
	\includegraphics[ width=2.0 in,height=2.0in]{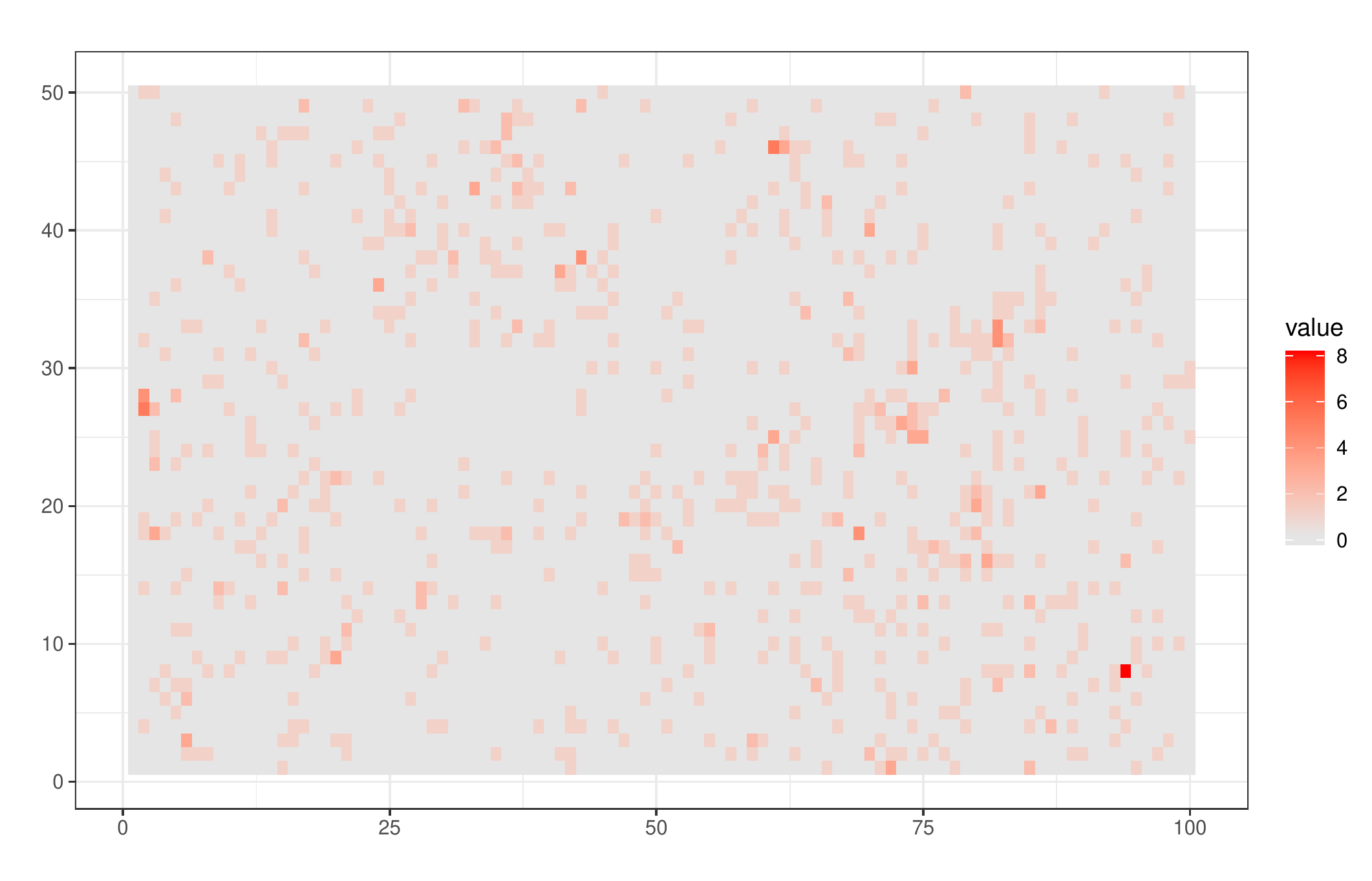} &
	\includegraphics[width=2.0 in,height=2.0in]{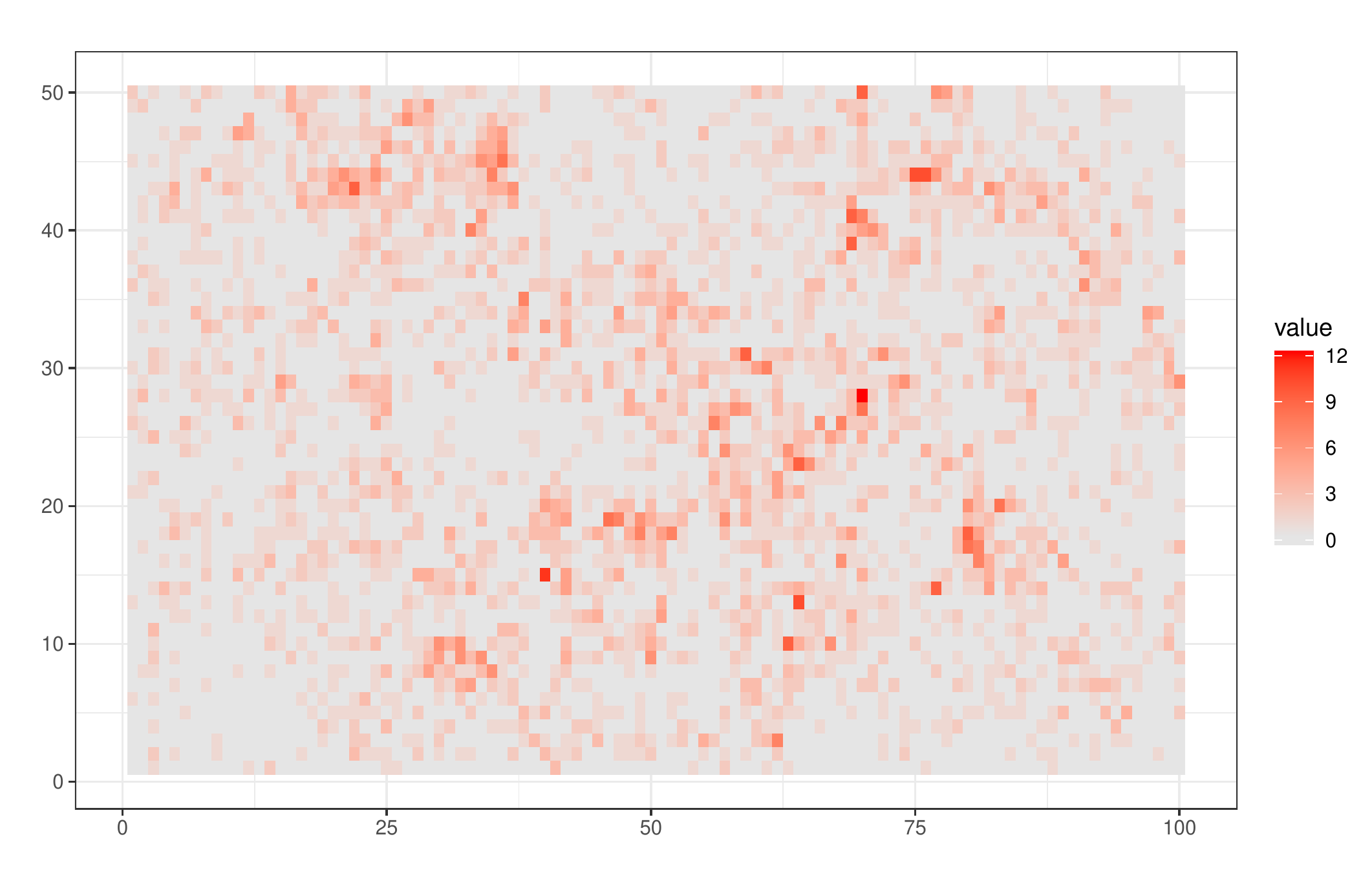} &
	\includegraphics[ width=2.0 in,height=2.0in]{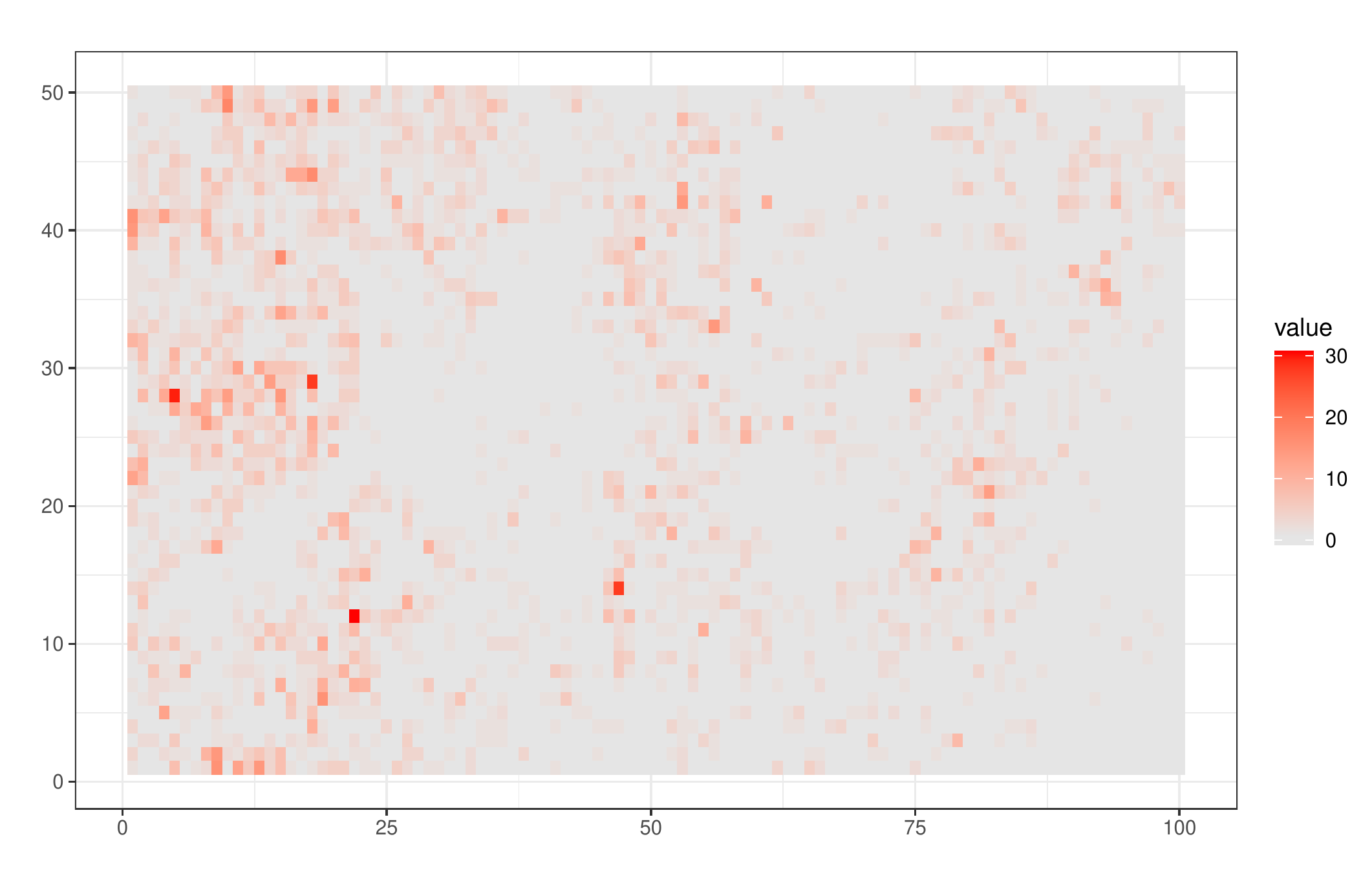} \\ [-0.25in]
	\includegraphics[width=2.0 in,height=2.0in]{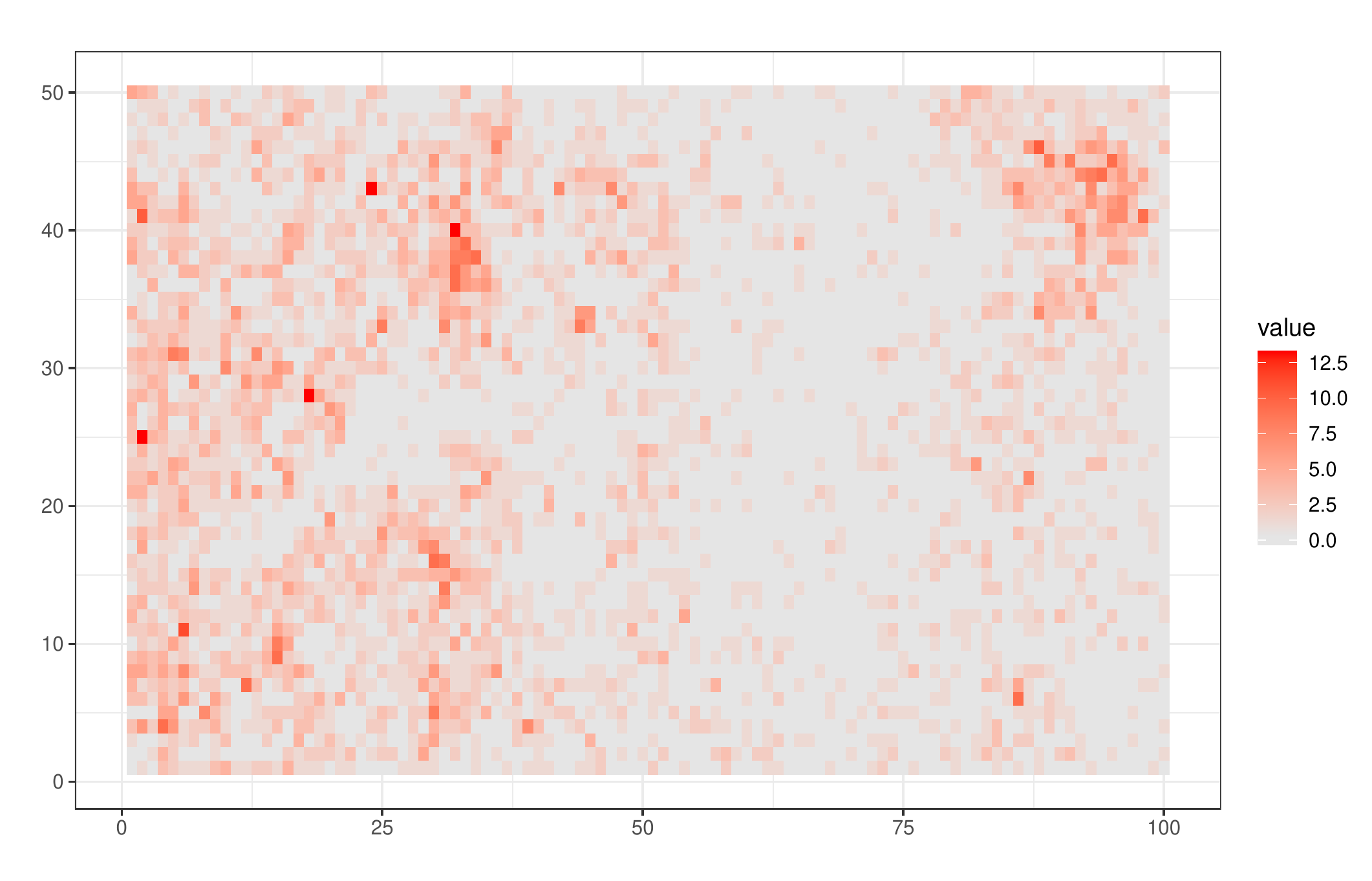} &
	\includegraphics[ width=2.0 in,height=2.0in]{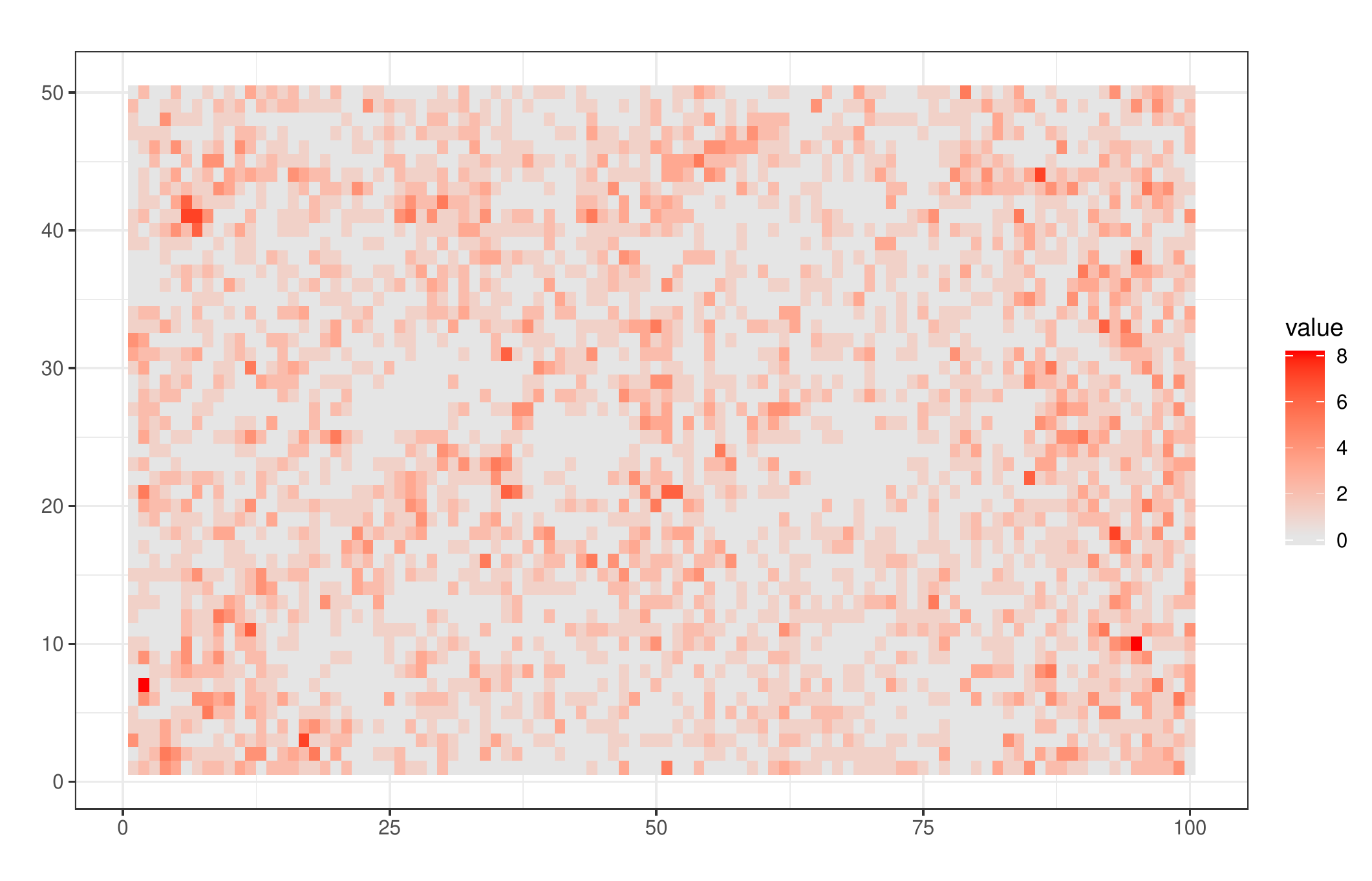} &
	\includegraphics[width=2.0 in,height=2.0in]{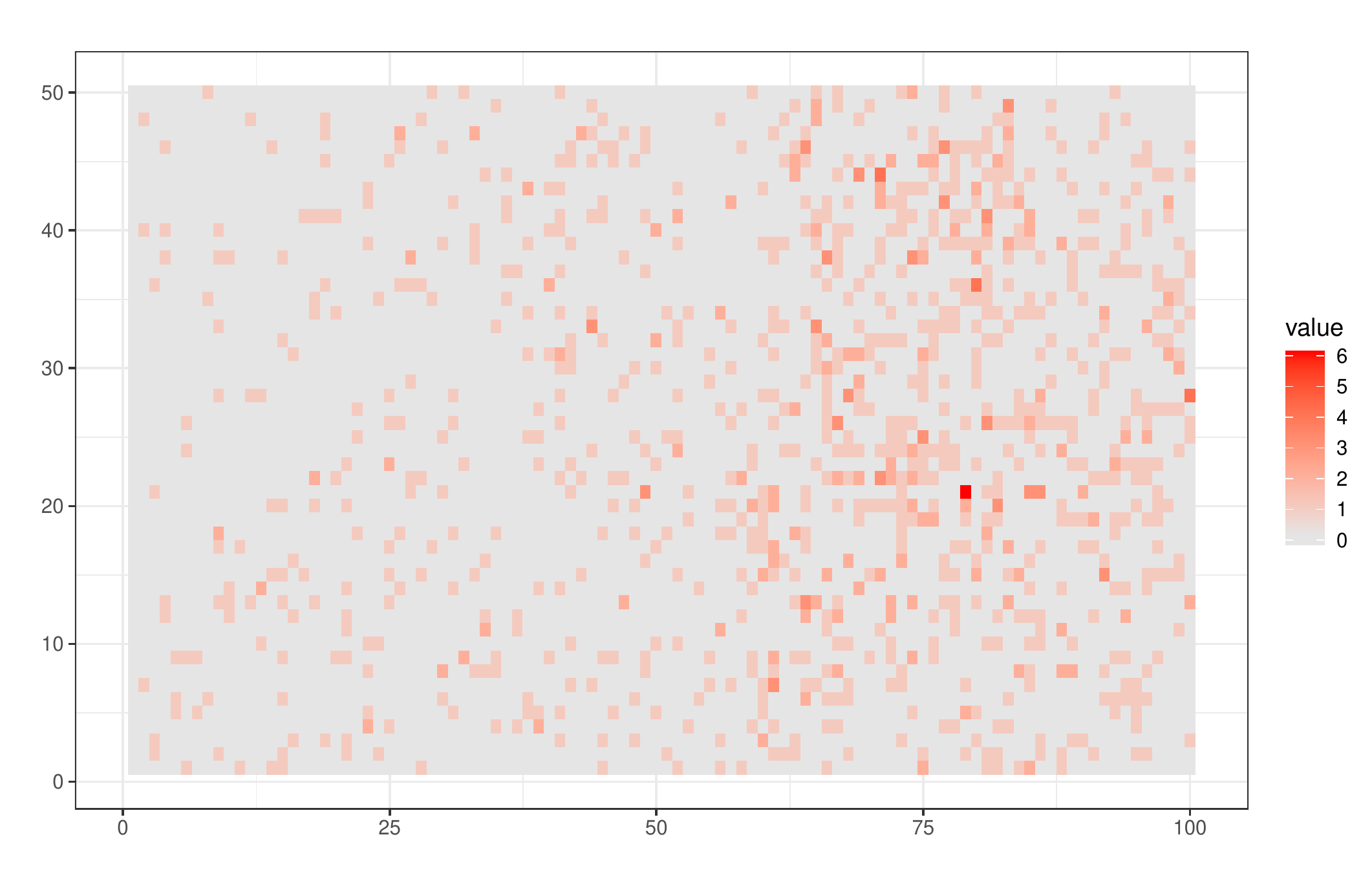} \\
 \end{tabular}
	
	\label{fig:covariate_map}
\end{figure}

The combination of the covariate model and the resolution with the smallest DIC value and the largest LPML value is 
the model with $(\beta_1,\beta_3,\beta_4,\beta_5)$ plus the $10 m \times 10 m$ resolution. The posterior means, the posterior standard deviations and the 95\% HPD intervals 
under this best combined covariate model and resolution are reported 
in Table \ref{table:BCI_estimates}. Figure \ref{fig:covariate_map} shows the plots of 6 different species in the $10 m \times 10 m$ pixel resolution.

\begin{table}[h]
\center
\caption{Posterior Estimation Results of BCI Data}
\label{table:BCI_estimates}
	\begin{tabular}{cccc}
	\hline
	Parameter& Posterior Mean& Posterior SD& 95\%HPD interval \\
	\hline
	$\beta_1$&0.2986 & 0.0275&(0.2415,0.3511)\\
	$\beta_3$& 0.1083&  0.0037&(0.1013,0.1156)\\
    $\beta_4$ &0.2305&  0.0064&(0.2180,0.2420)\\
    $\beta_5$ & 0.2825 & 0.0122 &(0.2595,0.3053)\\
 $\lambda_0$  &1200.4259 &25.3233&(1157.8434,1254.6554)\\
 \hline
	\end{tabular}
\end{table}

From Table \ref{table:BCI_estimates}, we see that (i) the \textit{B.pendula} trees seem to be attracted to \textit{Eugenia oerstediana}, 
\textit{Piper cordulatum}, \textit{Protium panamense}, and \textit{Sorocea affinis}; and (ii) \textit{Eugenia nesiotica} and \textit{Talisia croatii} have no effects on \textit{B.pendula} trees.

\section{Discussion}\label{sec:discuss}

In this paper, we have developed the Bayesian parameter estimation and model selection in Poisson point process models in regression context. Our simulation results indicate that our proposed method is more likely to select the correct model and obtain the estimators that are closer to the true values if we had fit the true model. Like most model selection methods, our proposed method needs calculating DIC and LPML for all candidate models. Developing a computational algorithm as in \cite{chen2008bayesian} is an important future work. In addition, extending our proposed criterion to a more complicated spatial point process like Gibbs point process \cite{moller2007modern} should be a natural extension of our work. Furthermore, another future work is to investigate the theoretical properties of the proposed procedure like \cite{li2017deviance}. Moreover, the development of the predictive distribution-based criterion like WAIC \cite{vehtari2017practical} or the L-measure \cite{ibrahim2001criterion} for 
spatial point process is another interesting future work.

\section*{Acknowledgement}
Dr. Chen's research was partially supported by NIH grants \#GM70335 and  \#P01CA142538. Dr. Hu's research was supported by Dean's office of College of Liberal Arts and Sciences in University of Connecticut.

\bibliographystyle{apacite}
\bibliography{Model_Selection_Point_Process_arxiv}

\end{document}